\documentstyle [12pt,amssymb,english,epsf] {article}
\begin {document}
\large {
\vspace * {6.0cm}
\begin {center}
{ \Large
O.V.Bulekov, N.D.Galanina, V.S.Demidov, \\ A.L.Endalov, I.L.
Kiselevich, M.A.Martem'yanov, \\ V.I.Mikhailichenko, V.A.Okorokov,
A.K.Ponosov, \\ F.M.Sergeev, N.A.Khaldeeva}

\vspace {2.0cm}

{ \bf {THE $ \Lambda K $ SYSTEM PRODUCTION IN THE HADRON INTERACTIONS }}
\end {center}

\vspace {10.0cm}

\begin {center}
Moscow 1998 \\
\end {center}

\clearpage
\baselineskip 25pt
\normalsize
UDK 539.1 \hspace*{12cm} ?-16 \\ \\
THE $ \Lambda K $ SYSTEM PRODUCTION IN THE HADRON
INTERACTIONS. \\ \\ O.V.Bulekov, N.D.Galanina, V.S.Demidov,
A.L.Endalov, I.L.Ki\-se\-le\-vich, M.A.Mar\-tem'ya\-nov,
V.I.Mikhailichenko, V.A.Okorokov, A.K.Po\-no\-sov, F.M.Ser\-ge\-ev,
N.A.Khaldeeva \\

The A-dependence is observed in $x_F$-distributions for the $\Lambda K^0$
system pro\-du\-ced with the small transverse momentum in the neutron-nucleus
interactions. For the $\Lambda$  hyperons  similar dependence isn't seen.
The result is interpreted as an effect from intermediate excitative nucleon
state, which decays into strange particles. Such interpretation is
confirmed  ex\-pe\-ri\-men\-tal data on $\Lambda K$ pair production in the
pion-nucleon interactions. \\ \\
Fig. - 8, References - 9
\vspace {1cm}

\copyright Institute of Theoretical and Experimental Physics, 1998
\vspace {1cm}
\begin {flushleft}
$\underline {~~~~~~~~~~~~~~~~~~~~~~~~~~~~} $ \\
$^{+)}$ Moscow state Engineering-Physical Institute
 (Technical University) \\
\end {flushleft}
\clearpage
\large
\vspace * {3 cm}

\begin {center}
I
\end {center}

The problems, connected to influence of the nuclear surroundings on
production of hadrons and the nuclear transparency (or in the modern
quantum mechanics - color shield and color transparency) is of great
attention of the scientists in the last years [1-3]. The characteristics of
producing leading particles are defined directly by evolution of quark
system. The leader represents some kind of "transformation" of a primary
particle, which characteristics reflect the history of hadron passage
through target. So, the cross section evolution of the forming hadron
can be possible to look on increasing (decreasing) of the secondary leading
particles beam intensity of various nuclei in comparison with the
hadron-nucleus interaction at appropriate kinematic intervals. The
research of the leading particles properties, distinguished on quarks
composition from the primary state, is of great interest. Avaricious
experimental data demonstrates, that the behaviour of such particles at
passage through nucleus differs from behaviour of the "kept" leaders. The
change of passing ability (transparency) of nucleus as well as energy
dependence is appeared to be brightly shown at initial enegries in the order
10 Gev already.

The characteristic, transmitted "optical" properties of a nucleus in relation
to hadron passage, is the effective number of nucleons, which represents the
ratio of differential sections in the appropriate ranges of Feynman's variable
$x_{F}$ at reactions on nuclei and free nucleons:
\begin{center}
$N_{eff}= \frac{\textstyle (\textstyle d \sigma / \textstyle då_{F})_
{\textstyle A}}
{\textstyle (\textstyle d \sigma / \textstyle då_{F})_{\textstyle N}}.$
\end{center}
The experimental data on inclusive production of pions and nucleons at initial
energy of 100 GeV are in the consent with calculations on dual string model -
the change of a nucleus passing ability for particles with different values of
Feynman's variable is observed. However, there is a reference for
$\Lambda$-hyperons on some divergence of experimental data with calculations
- probably, the passing ability of nucleus is identical for particles with
different values of the Feynman's variable [1].

At study of leading $\Lambda$-hypeorns by neutrons with energy of 4 - 10 GeV,
we showed, that the cross sections relation of hyperon's production on nuclei
of carbon and lead don't depend from Feynman's variable [4].

In the present work, we investigated production of the leading pairs
$\Lambda K^0$ by neutrons on carbon and copper nuclei in an angular range
$\Theta < 8.5^0$ in relation of beam and supporting only neutral particles.
The experiment was performed at KAON spectrometer on neutron beam of
ITEP's proton synchrotron [5]. As a target was used graphite by thickness
of 6 cm and copper by thickness of 4 cm. $\Lambda$-hyperons and
$K^0$-mesons were registered by magnetic spectrometer KAON using wire spark
chambers and system of the scintillation counters. The target was positioned
on a beam, between the scintillation counters, connected in anticoincidence
with other counters. Thus, the events accompanying by neutral strange particles
were excluded, that has allowed to allocate half inclusive reaction:
\begin{equation}
n + ( C, Cu )\rightarrow \Lambda + K^0 + X^{0}.
\end{equation}
The decays of $\Lambda$ and $K^0$-particles, produced in target, were
detected by spectrometer on base length of 40 cm. The decaying volume was
filled by helium. The start of installation was carried out by the trigger from
the circuits of fast electronic logic, using signals of scintillation counters.
The trajectories of the charged particles were registered by groups of wire spark
chambers.

The accumulated statistics was involved 349 pairs $\Lambda K^0$, including 215
on carbon target, 134 - on copper target. The registration efficiency of
$\Lambda K^0$ pair was calculated by the software package GEANT.

The amendments on interaction, producing together with pair $\Lambda K^0$ with
neutrons and $\pi^0$-mesons with target nuclei, were entered.

At realization of the experiment modeling, it was considered, that the reaction
can be carried out as well on a nucleus neutron as on a proton. Therefore,
the he contributions from four reactions were taken into account:
\begin{equation}
 n + n \to  n + \Lambda + K^{0},
\end{equation}
\begin{equation}
 n + n \to  n + \Lambda + K^0 + \pi^{0},
\end{equation}
\begin{equation}
 n + p \to  p + \Lambda + K^{0},
\end{equation}
\begin{equation}
 n + p \to  p + \Lambda + K^0 + \pi^{0}.
\end{equation}
The relative contribution of reactions on proton and neutron was considered at
conformity with the number of protons and neutrons in nuclei of carbon and
carbon. The contributions of three and four particles reactions were taken
into account proportionally their cross section taken from the world data.
Thus, the cross section of connected isotopically $pp$-reactions were used
for reactions (2) and (3). The used modeling procedure has shown, that
process take place on the neutrons of nuclei mainly. The reactions, proceeding
on protons of a nucleus, are suppressed by inclusion of the anticoincidence
counter. Nevertheless,  the reaction on protons of a nucleus take place,
because the secondary protons can stop in target or don't hit in the
counter of anticoincidence. The contribution from a four particles reaction
is less than the contribution from a three particles, because electron-positron
pairs, produced at conversion of $\gamma$-rays from $\pi^0$-meson decay, can be
a cause of counter's operation.

The dependence of distribution on $x_{F}$ from the mass number of nucleus-target,
characterized for the leading mesons, isn't observed for leading $\Lambda$-
hyperons. The distribution on Feynman's variable for $\Lambda K^0$-system
produced on in a copper target is presented in Fig. 1, in Fig. 2 - for a
carbon target. The relation R of differential cross section of $\Lambda K^0$
pairs production, characterizing the passing ability of nuclei, is
appeared to depend from Feynman's variable (Fig. 3). This result can be
understood, if heavy isobar is assumed to produced originally and decayed
on a strange particles after that. On the other hands, the $\Lambda K^0$-
system at pasage through a nucleus behaves as one unstrange meson but nor
as two strange particles.

The limited statistics does not allow to analyze spectra of effective mass of
$\Lambda K^0$ system at detail. However, it should be noted, that spectrums of
effective mass, obtained on a carbon and copper target, are different, i. e.
the nuclear surroundings influences on kinematical characteristics of formed
particles even under the condition of cascade processes suppression.

The spectrums of the effective masses were approximated by the sum of
Breit-Wigner distributions and phase volume.  The results of approximation
are submitted in figures by a continuous line and given in Table 1. The
essential increase of statistical accuracy is necessary for more detailed
study of the effective mass spectrums.
\clearpage
\begin{center}
\tabcolsep 0.5cm
{\bf ' ble 1.}
The approximation results of the invariant mass spectrums of
$\Lambda K^0$-system for carbon and lead targets. \\
\vspace*{0.5cm}
\begin{tabular}{|c|c|c|p{3.0cm}|}
\hline
\multicolumn{1}{|c|}{Peremeter}
& \multicolumn{1}{|c|}{Target C}
& \multicolumn{1}{|c|}{Target Cu} \\
\hline
$M_0, ŒeV/c^2 $ & $1797\pm 6 $ & $1743\pm 13$  \\
\hline
$ƒ, MeV/c^2$ & $ 176\pm 25 $& $72\pm 21$ \\
\hline
$\chi^2$/degrees of freedom & 9.7/7  &  6.3/8  \\
\hline
\end{tabular}
\end{center}
\vspace{0.5cm}

It is necessary to essentually increase the statistical accuracy for
quantitative results. At the same time the confirmation assumption of
an intermediate isobar role can serve the data, obtained at the
analysis of the secondary particles effective mass spectrums in
$\pi^+p$-interactions with production of the strange particles.

\begin {center}
II
\end {center}

The resonance production in $\pi ^ {+}p$-interactions at an initial momentum
of 4,23 GeV/c with strange particles production was investigated by us on a
material of 2-meter by hydrogen camber of ITEP at three channels using rather
large section:
\begin{equation}
\pi^+ + p \to \Lambda + K^+ + \pi^+,
\end{equation}
\begin{equation}
\pi^+ + p \to \Lambda + K^+ + \pi^+ + \pi^0,
\end{equation}
\begin{equation}
\pi^+ + p \to \Lambda + K^0 + \pi^+ + \pi^+ .
\end{equation}

Due to analysis the events were used, where beam momentum, founded from the
balance of energy, differed from average no more than 75 MeV/c - experimentally
determined beam dispersion. It is made to elimination of the badly balanced
events, which presence can deform mass spectrums that, will have an effect
on adjustment results.

The hypotheses of resonance production, number of events, obtained on the full
statistics, and cross sections of channels, performed at simultaneous adjustment
by multiparametrical function, containing resonance amplitudes generated by a
method of Monte-Carlo, their mutual reflections and phase volumes to
experimental distributions on effective mass, are presented in Tables 2-4.
Due to this procedure the effects of resonace reflections for all mass
spectrums were taken into account.

The minimization was carried out by the method of least squares with help of
program MINUIT [7]. The errors in cross sections are statistical. Effective
masses and widths of well known resonances
$Š^*$(892), $\Sigma$(1385) and $\rho$ were taken from the table of elementary
particles, and an appropriate parameters of isobar and resonances $\Sigma$(1560),
$\Sigma$(1670) and $\Sigma$(1849) were defined by minimization process.

The value of the invariant mass of isobar positioned  in the region closely to
1.7 GeV/c$^2$, width - about 100 MeV/c$^2$. Due to this region a little close
located isobars are known, that does not given an opportunity to identify
unequivocally the discovered resonance. Most likely, it should be identified with
$N^*$(1710) 1/2$^+$-isobar having the  high probability of decay into
$\Lambda$ and K. Due to the benefit of such assumption the close location to
isotropic angular distribution of $\Lambda$-hyperons according  direction of
$\Lambda K$ momentum in thier system of rest is testify.
The given results testify about the large contribution to cross section channels
with resonance produtcion in $\pi^+p$-interactions at 4.23 GeV/c. More than
80\% of the investigated reactions take place through resonance production.

The experimental distribution (histograms) on effective masses of  $\Lambda K$
together with curves, obtained as fitting result, are presented in Fig. 6-8 for
an illustration of the fitting quality.

\begin{center}
\tabcolsep 0.5cm
{\bf Table 2.}
Cross section of resonances production for the channel $\Lambda K^+ \pi^+$\\
\vspace*{0.5cm}
\begin{tabular}{|c|c|c|p{3.0cm}|}
\hline
\multicolumn{1}{|c|}{Number of events}
& \multicolumn{1}{|c|}{Hypothesis}
& \multicolumn{1}{|c|}{Cross section, ¬cb}
\\ \hline
    &$\Lambda K^+ \pi^+$  &  26,6 $\pm$ 1,6 \\
387 &$\Sigma(1385)K^+$    &  21,2 $\pm$ 1,4 \\
    &$\Sigma(1670)K^+$    &  10,6 $\pm$ 1,0 \\
    &$N^*(1710)\pi^+$     &  5,8 $\pm$ 0,8 \\
\hline
\end{tabular}
\end{center}
\vspace{0.5cm}
\begin{center}
\tabcolsep 0.5cm
\clearpage
{\bf Table 3.}
Cross section of resonances production for the channel $\Lambda K^+ \pi^+\pi^0$\\
\vspace*{0.5cm}
\begin{tabular}{|c|c|c|p{3.0cm}|}
\hline
\multicolumn{1}{|c|}{Number of events}
& \multicolumn{1}{|c|}{Hypothesis}
& \multicolumn{1}{|c|}{Cross section, mcb}
\\ \hline
     &$\Lambda K^+ \pi^+\pi^0$    &    15,5 $\pm$ 1,3  \\
     &$\Sigma(1385)K^+\pi$        &    52,9 $\pm$ 2,4  \\
     &$\Sigma(1560)K^+\pi$        &    11,7 $\pm$ 1,1  \\
     &$\Sigma(1670)K^+\pi$        &    14,4 $\pm$ 1,3  \\
841  &$\Lambda K^*(892) \pi^+$    &    10,2 $\pm$ 1,1  \\
     &$N^*(1710)\pi^+\pi^0$       &    17,2 $\pm$ 1,4  \\
     &$\Lambda K^+ \rho^+$        &    10,4 $\pm$ 1,1  \\
     &$N^*(1710)\rho^+   $        &     0 \\
     &$\Sigma(1385)K^*(892)$      &    15,4 $\pm$ 1,3  \\
\hline
\end{tabular}
\end{center}
\vspace{0.5cm}
\begin{center}
\tabcolsep 0.5cm
{\bf Table 4.}
Cross section of resonances production for the channel $\Lambda K^0 \pi^+\pi^+$\\
\vspace*{0.5cm}
\begin{tabular}{|c|c|c|p{3.0cm}|}
\hline
\multicolumn{1}{|c|}{Number of events}
& \multicolumn{1}{|c|}{Hyphothesis}
& \multicolumn{1}{|c|}{Cross section, mcb}
\\ \hline
    &$\Lambda K^0 \pi^+\pi^+$    &   0 \\
    &$\Sigma(1385)K^0\pi^+$      &   30,8 $\pm$ 1,8  \\
    &$\Lambda K^*(892) \pi^+$    &    7,7 $\pm$ 0,9  \\
563 &$N^*(1710)\pi^+\pi^+$       &   20,7 $\pm$ 1,5  \\
    &$\Sigma(1385)K^*$           &   14,7 $\pm$ 1,2  \\
    &$\Sigma(1670)K^0\pi^+$      &    4,1 $\pm$ 0,6  \\
    &$\Sigma(1849)K^0\pi^+$      &    5,9 $\pm$ 0,8  \\
    &$\Lambda K^*(1430) \pi^+$   &   10,2 $\pm$ 1,0  \\
\hline
\end{tabular}
\end{center}
\vspace{0.5cm}

The data about heavy isobar, decaying into strange particles were obtained from the
analysis of the effective mass spectra in reaction (6) in works, performed on the
chamber of Alvaretz [8,9].

The basic results of the present work are following.\\

1.  The dependence of Feynman's variable distribution from mass number of
nucleus target is observed in reaction of $\Lambda K^0$ production by neutrons
on carbon and copper nuclei. Qualitatively, this dependence is coordinated to
calculation of B.Z. Kopeliovich on the dual string model for inclusive hadrons
production. \\
2. The observable dependence can be explained by production of intermediate
isobar, decaying on a strange particles. \\
3. This interpretation proves to be true by experimental data, obtained by us
from the analysis of effective masses spectra in $\pi^+p$-interactions with
generation of strange particles. More than 80\% of these reactions at an
initial moment of 4,2 Gev/c carry out through production of resonances. \\

The authors are grateful L.A. Kondratuk for discussions and useful councils.

The work is performed at financial support of RFFI (code of the project 97-02-17862).
\clearpage

\begin {center}
 References
\end {center}

1. B.Z.Kopeliovich, L.B.Litov, J.Nemchik. Effects of formation
time and colour transparency on the production of leading particles off
Nuclei. Preprint JINR, E2-90-344, Dubna, 1990.

2. Y.S.Golubeva, L.A.Kondratyuk, A.Bianconi, S.Boffi, M.Radici.
Nuc\-lear transparency in quasielastic A (e, e'p): Intranuclear cascade versus

3. L.A.Kondratyuk, Ye. A.Golubeva. Effects of nuclear medium in the formation
and production of hadronic resonances on nuclei. // Yad. Phys, 1998, Vol.61, p.951-
960.

4. O.V.Buekov et al. Characteristics of leading $\Lambda$-hyperons, produced by
neutrons on nuclei. M., Preprint ITEP, 1997, N4. // Yad. Phys., 1998, Vol.61, p.80-85.

5. A.N.Alexeev, V.M. Berezin, E.T.Bogdanov et al. Pair productiuon of $\Lambda K^0$ at
interaction of neutrons with nuclei. // Yad. Phys., 1991, Vol.54, p.1597-1604.

6. Y.D.Aleshin et al. Two-meter liquid hydrogen chamber of ITEP. // PTA, 1970, N.3,
p.100-102.

7. F.James and M.Roos, CERN Program Library D506, MINUIT - Function
Minimization and Error Analisis. 1988.

8. D.J.Crenell et al. $ \Lambda K $ enhancement at 1.7 GeV produced in $\pi p
\to \Lambda K \pi $ interactions at 6 GeV/c. // Phys. Rev. Lett., 1967, V.19,
P.1212-1215.

9. D.J.Crenell et al. Properties of the $N^*(1730)$. // Phys. Rev. Lett.,
1970, V.25, P.187-191.

\evensidemargin -10mm
\oddsidemargin -10mm

\newpage
\vspace{-2cm}
\begin{figure}[h]
\begin{center}
\mbox{ \epsfysize=14.0cm \epsffile{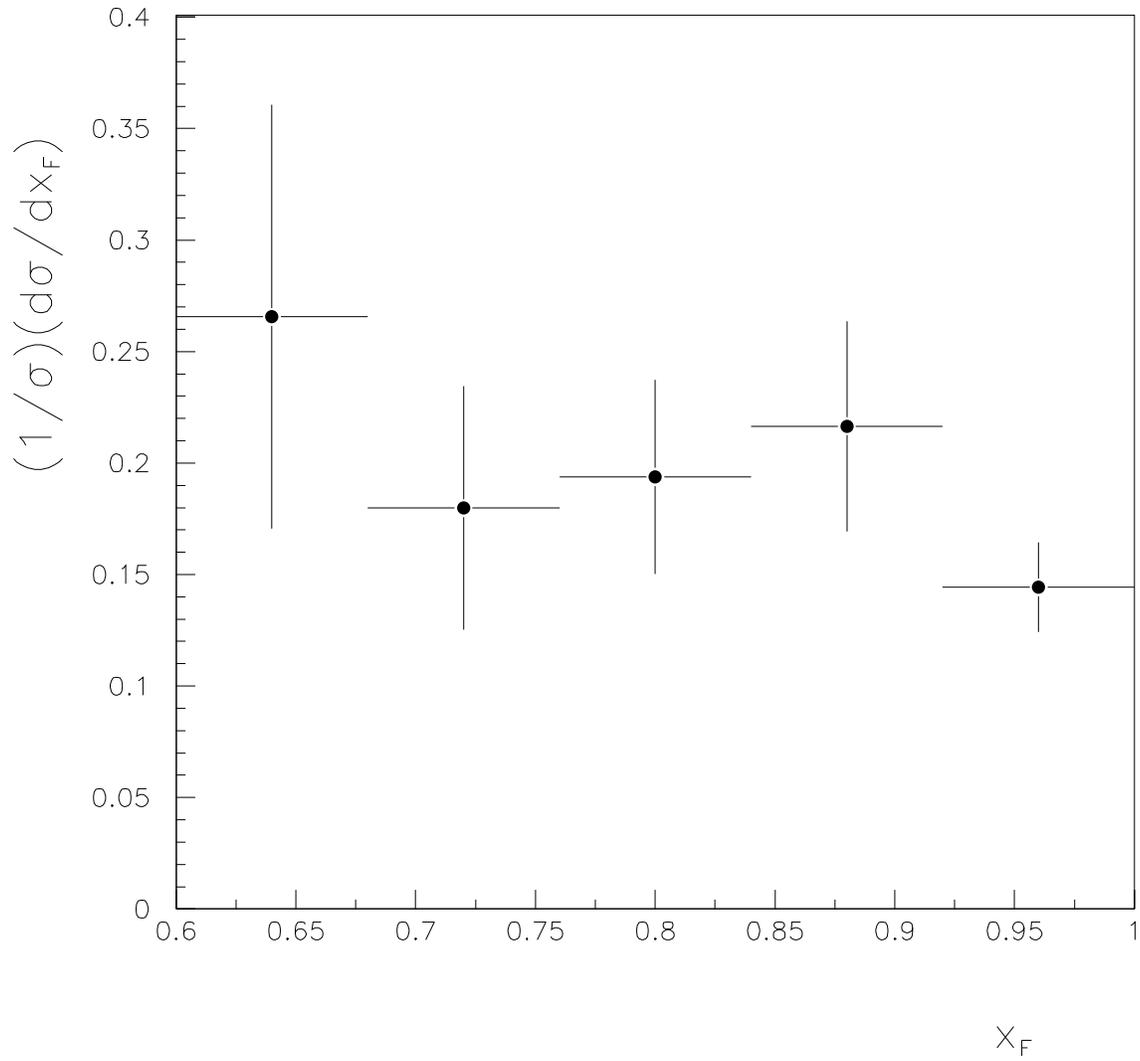} }
\caption{The distribution on Feynman's variable for
$\Lambda K^0$ system produced on a copper target.}
\end{center}
\end{figure}

\newpage
\vspace{-2cm}
\begin{figure}[h]
\begin{center}
\mbox{ \epsfysize=14.0cm \epsffile{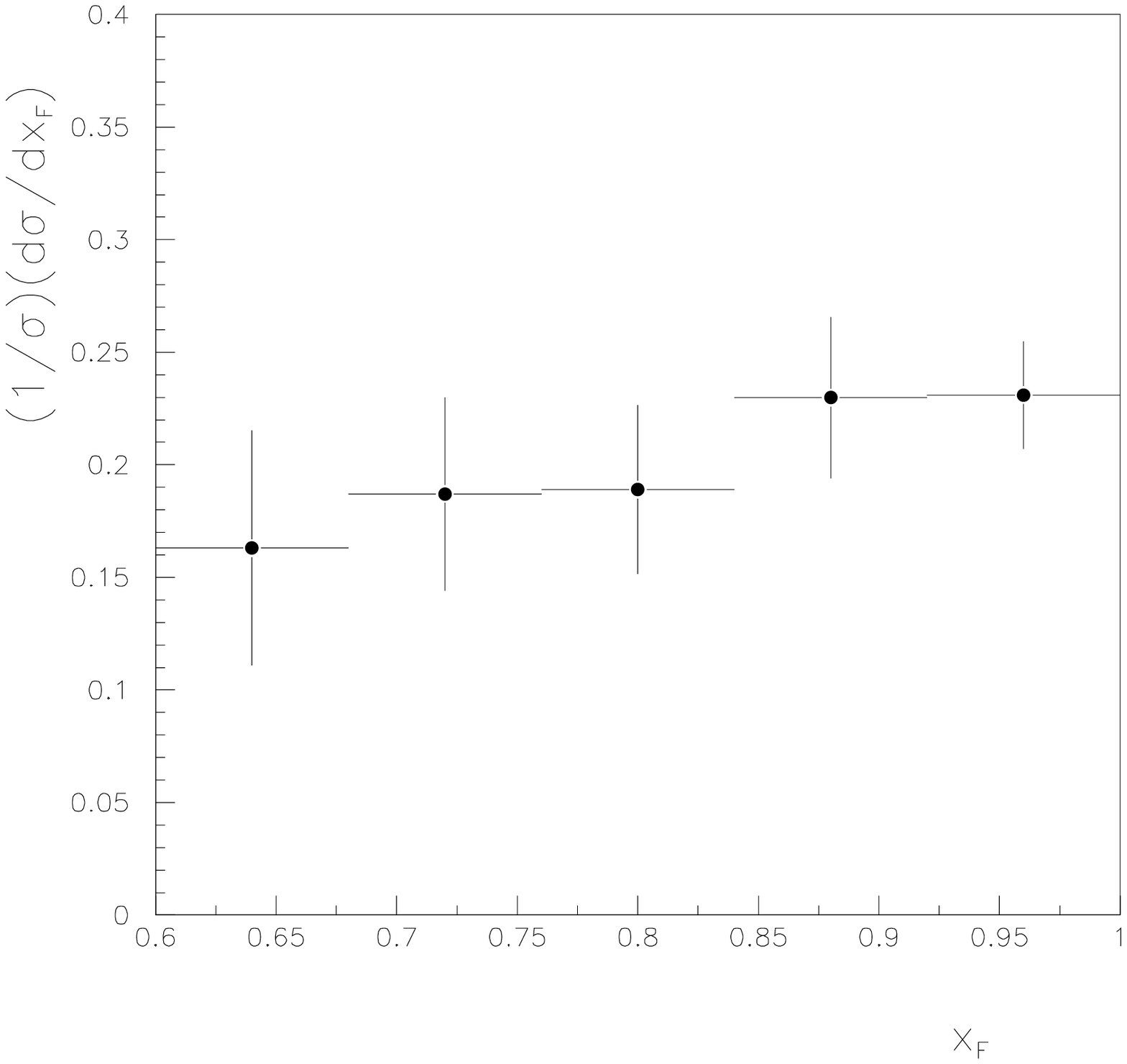} }
\caption{
The distribution on Feynman's variable for
$\Lambda K^0$ system produced on a carbon target. }
\end{center}
\end{figure}

\newpage
\vspace{-2cm}
\begin{figure}[h]
\begin{center}
\mbox{ \epsfysize=14.0cm \epsffile{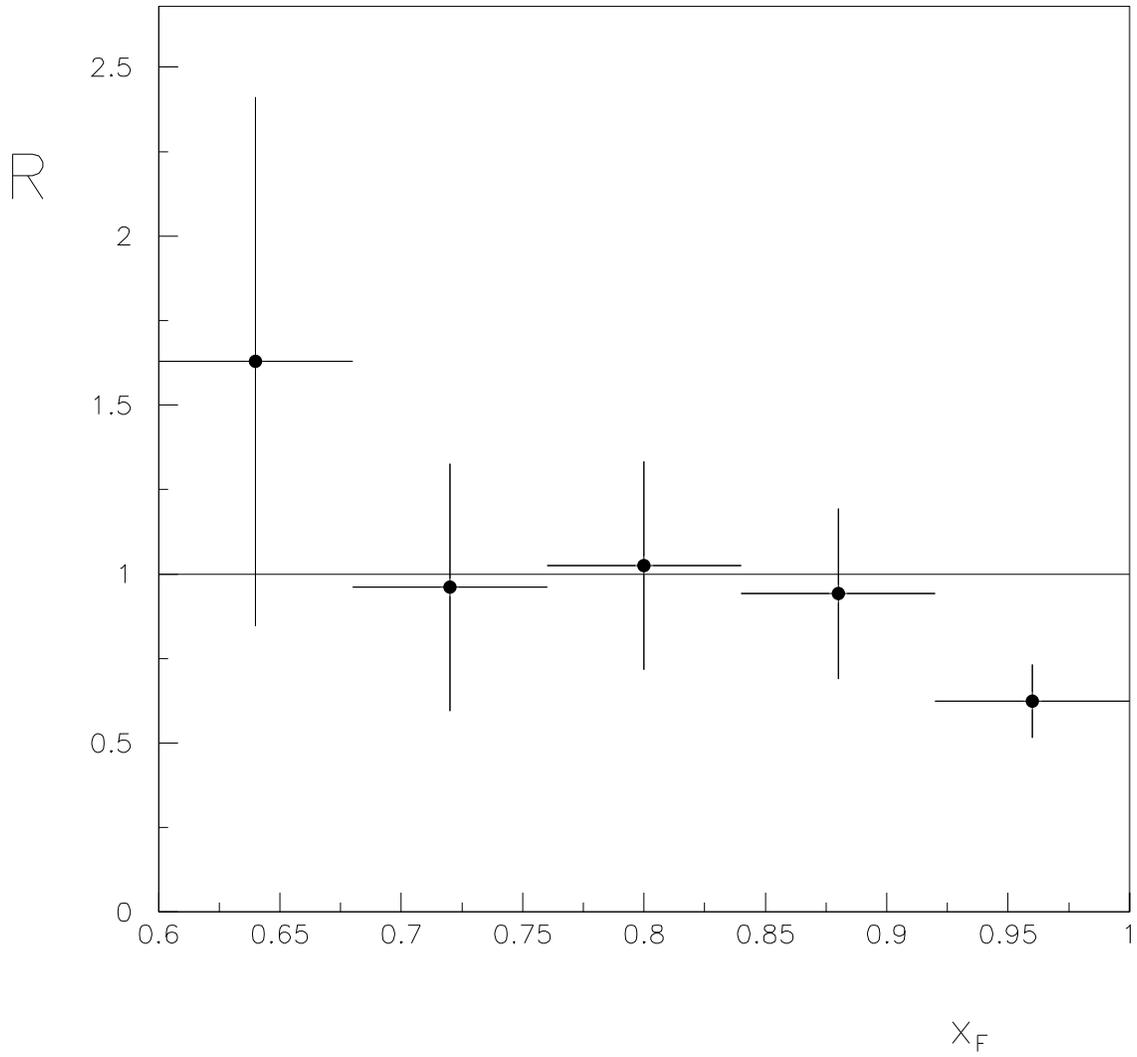} }
\caption{
The ratio of differential sections of pair production $\Lambda K^0$ on copper
and carbon targets depending on Feynman's variable. }
\end{center}
\end{figure}

\newpage
\vspace{-2cm}
\begin{figure}[h]
\begin{center}
\mbox{ \epsfysize=14.0cm \epsffile{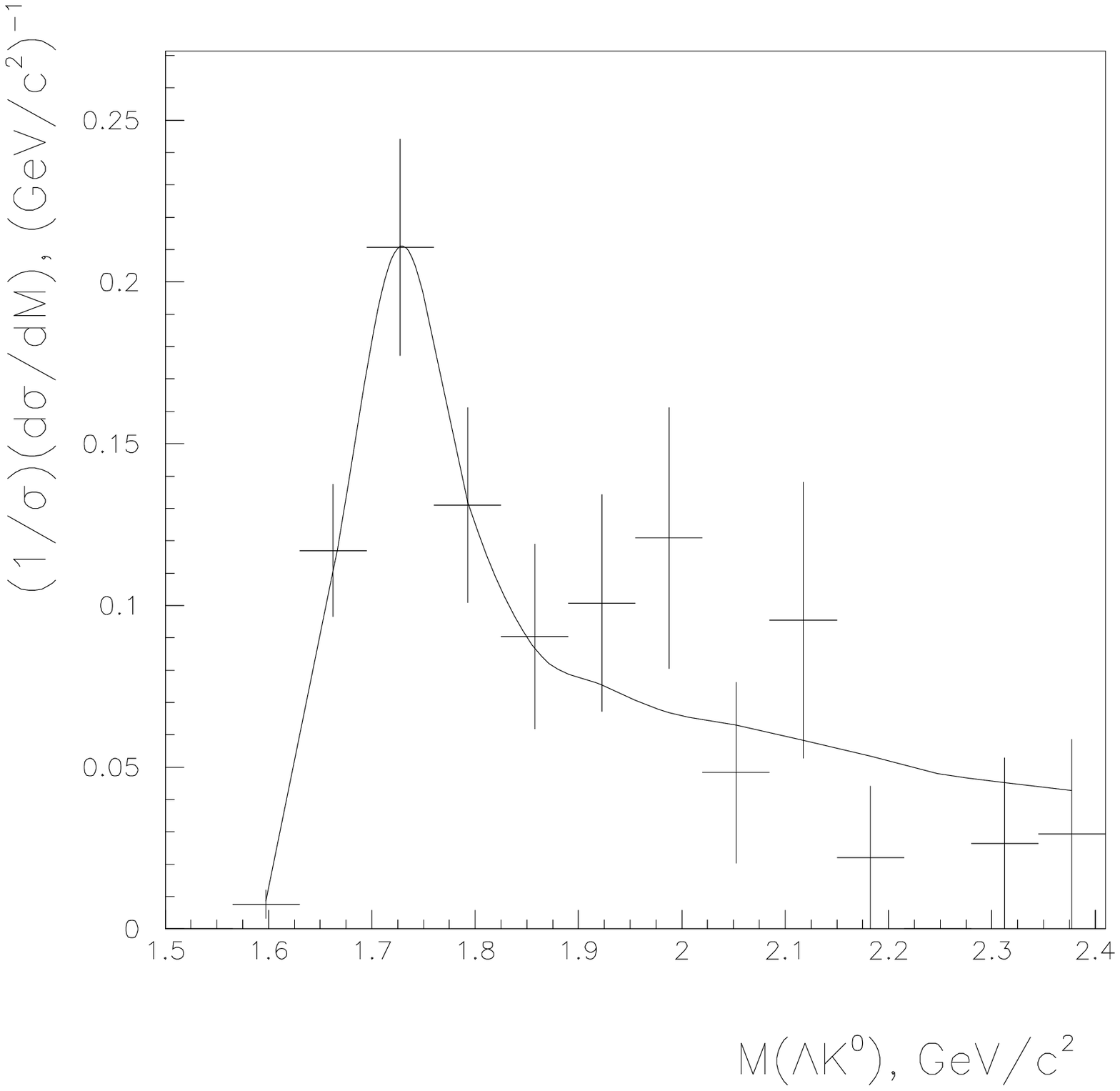} }
\caption{
The spectrum of effective masses of $\Lambda K^0$ system produced by neutrons
on copper target. }
\end{center}
\end{figure}

\newpage
\vspace{-2cm}
\begin{figure}[h]
\begin{center}
\mbox{ \epsfysize=14.0cm \epsffile{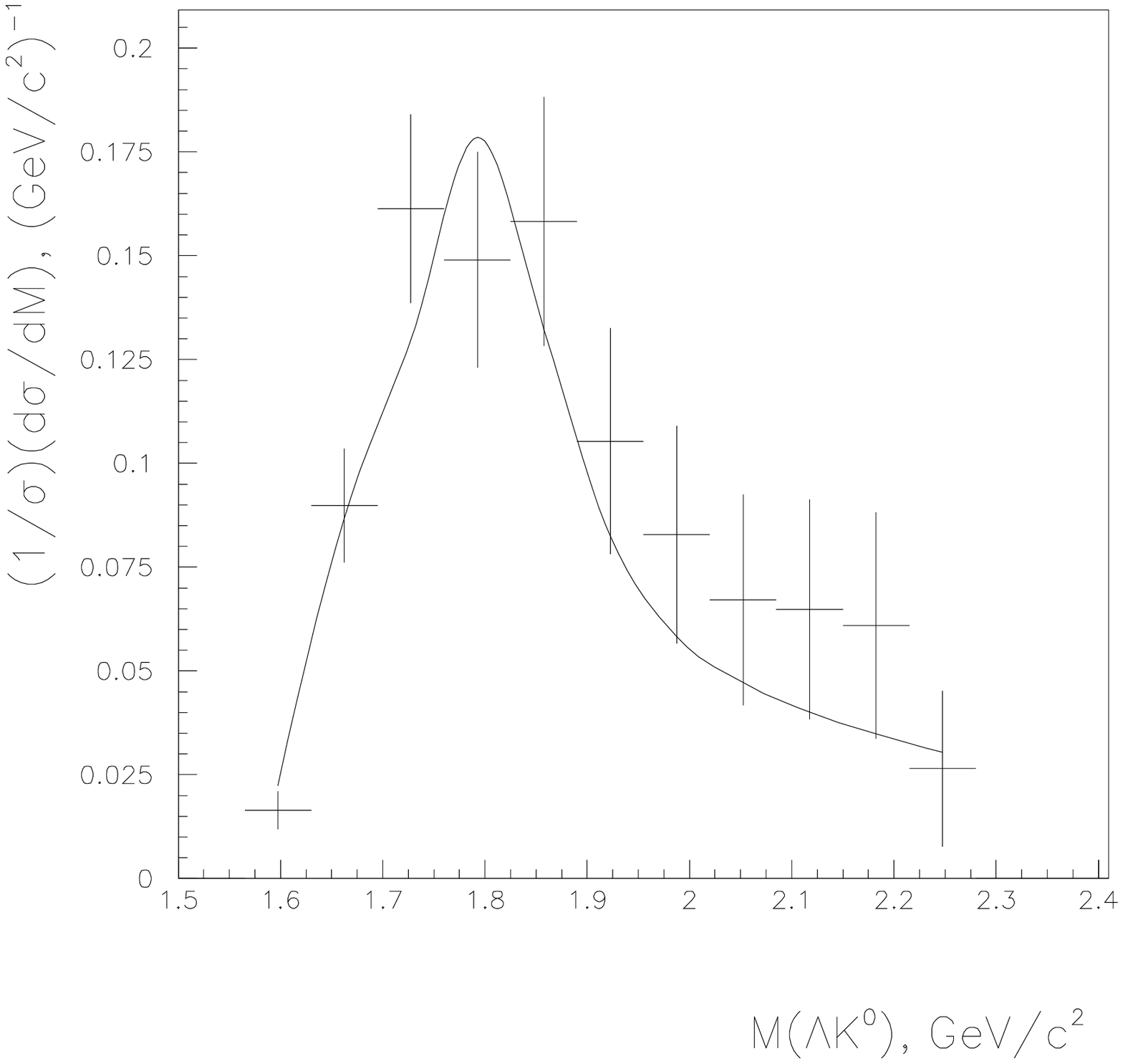} }
\caption{
The spectrum of effective masses of $\Lambda K^0$ system produced by neutrons
on carbon target. }
\end{center}
\end{figure}

\newpage
\vspace{-2cm}
\begin{figure}[h]
\begin{center}
\mbox{ \epsfysize=14.0cm \epsffile{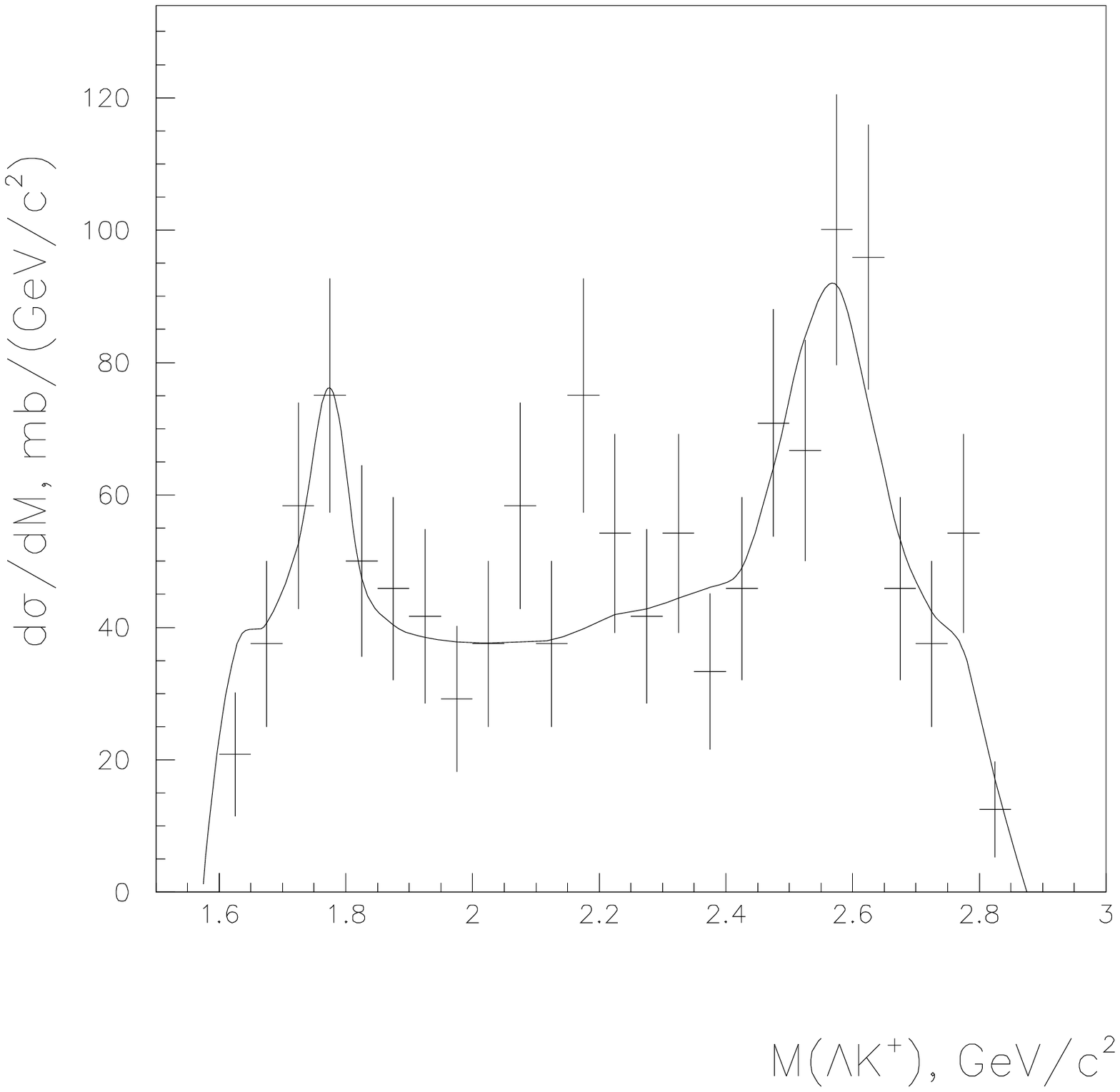} }
\caption{
The spectrum of effective masses of $\Lambda K^+$ system produced in reaction
(6). }
\end{center}
\end{figure}

\newpage
\vspace{-2cm}
\begin{figure}[h]
\begin{center}
\mbox{ \epsfysize=14.0cm \epsffile{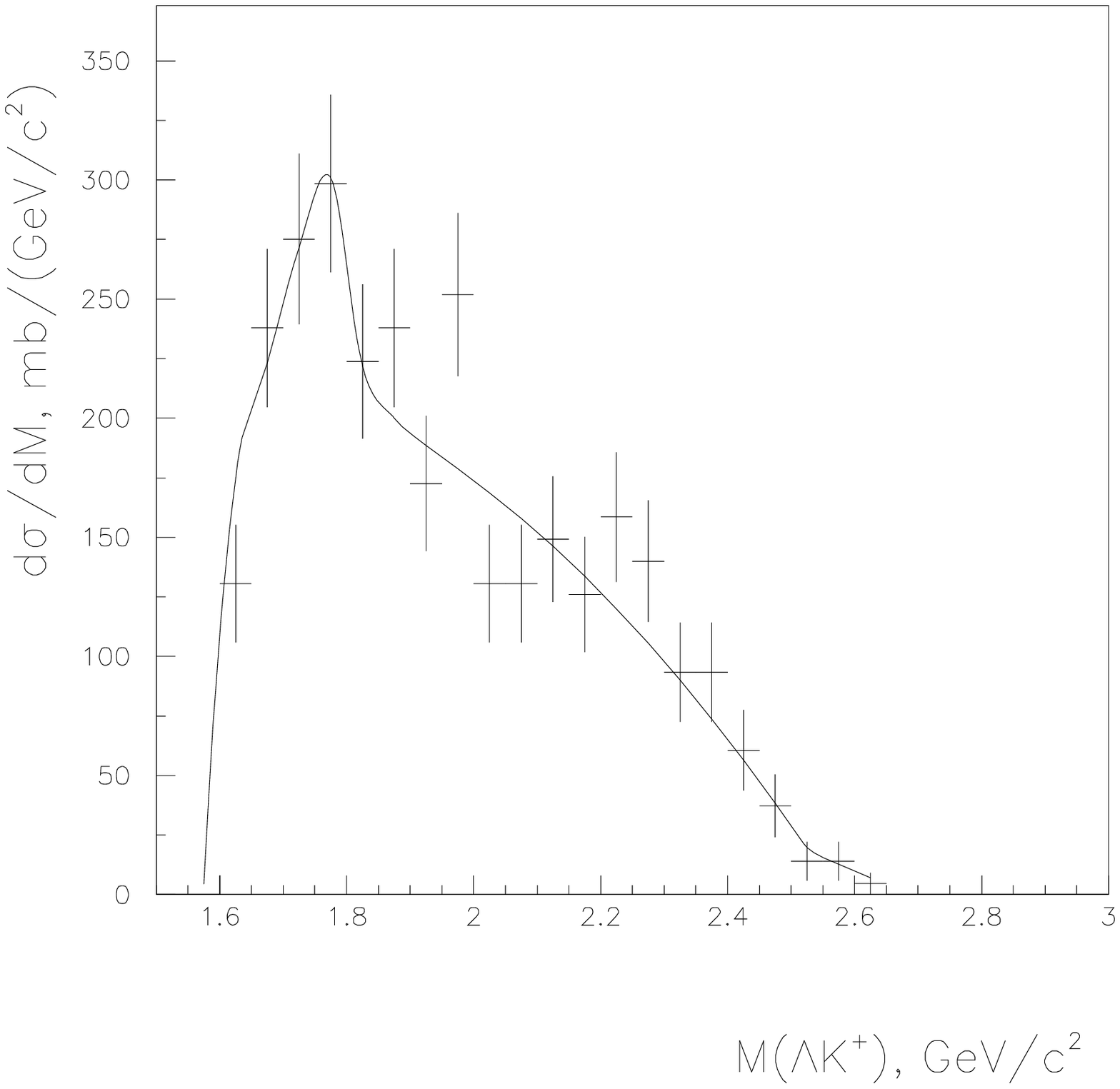} }
\caption{
The spectrum of effective masses of $\Lambda K^+$ system produced in reaction
(7).}
\end{center}
\end{figure}

\newpage
\vspace{-2cm}
\begin{figure}[h]
\begin{center}
\mbox{ \epsfysize=14.0cm \epsffile{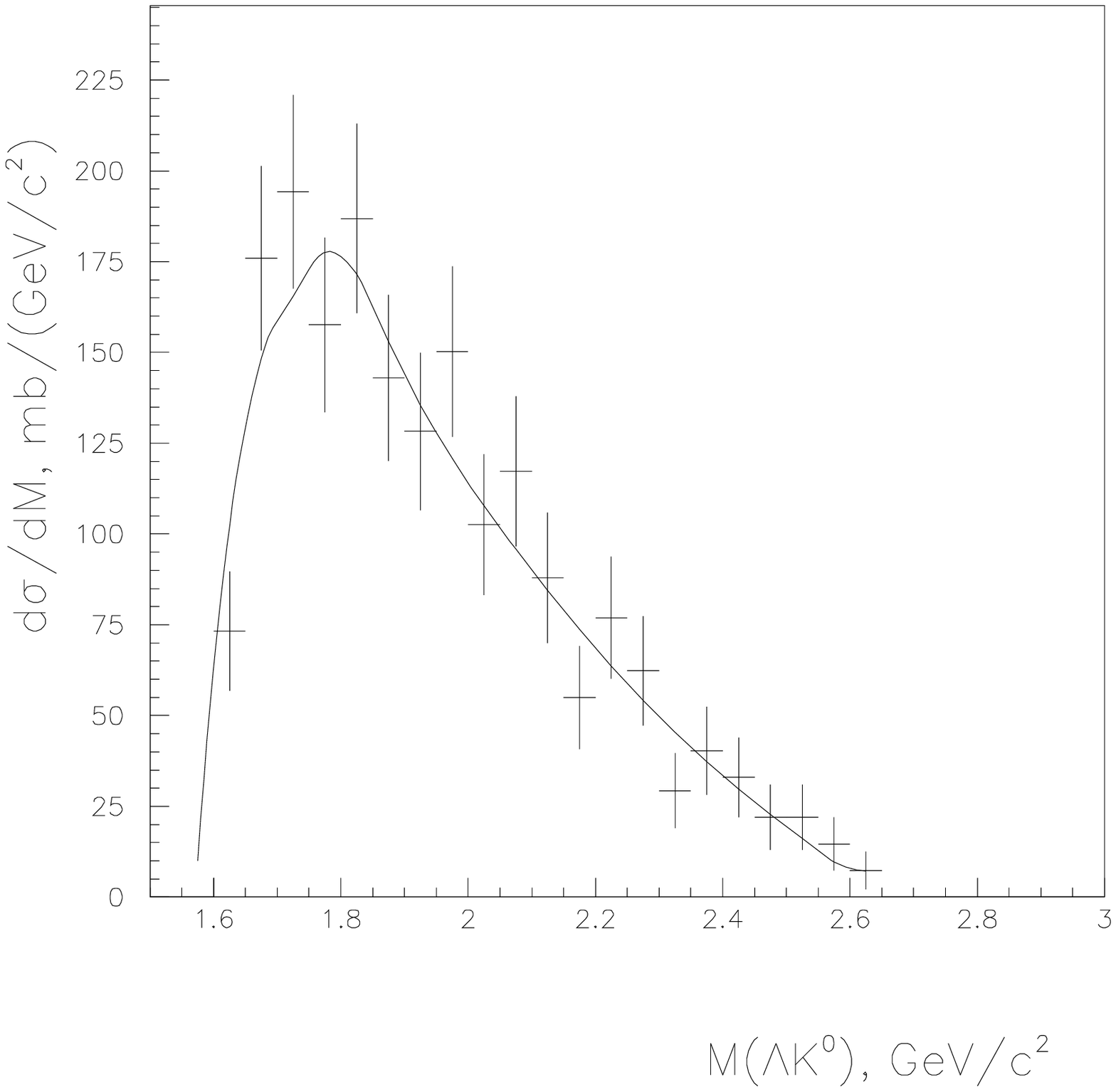} }
\caption{
The spectrum of effective masses of $\Lambda K^0$ system produced in reaction
(8).}
\end{center}
\end{figure}

\end {document}